\pdfoutput=1
\documentclass[11pt]{article}

\usepackage[]{EACL2023}

\usepackage{times}
\usepackage{latexsym}

\usepackage{amsmath}
\usepackage{amsfonts}
\usepackage{amssymb}

\usepackage{float}

\usepackage{graphicx}
\usepackage{float}
\usepackage{multirow}

\usepackage{svg}
\usepackage{xcolor}

\usepackage{booktabs,subcaption,amsfonts,dcolumn}
\usepackage[flushleft]{threeparttable}

\usepackage[T1]{fontenc}
\usepackage[utf8]{inputenc}

\usepackage{microtype}

\usepackage{inconsolata}

\usepackage{collcell}
\makeatletter
\newcolumntype{G}{>{\collectcell\@gobble}c<{\endcollectcell}@{}}
\makeatother

\usepackage[normalem]{ulem}
\title{Language Agnostic Multilingual Information Retrieval \\with Contrastive Learning}

\author{Xiyang Hu$^{1}$, Xinchi Chen$^{2*}$, Peng Qi$^{2*}$, Deguang Kong$^{2}$, Kunlun Liu$^{2}$, \\ 
{\bf William Yang Wang$^{2}$, Zhiheng Huang$^{2}$} \\
  $^{1}$Carnegie Mellon University \\
  $^{2}$AWS AI Labs \\
  \texttt{xiyanghu@cmu.edu},
  \texttt{\{xcc,pengqi,kongdegu,kll\}@amazon.com} \\
  \texttt{wyw@amazon.com}, \texttt{zhiheng@amazon.com}
  }

\begin{document}

\maketitle
\begin{abstract}
Multilingual information retrieval (IR) is challenging since annotated training data is costly to obtain in many languages. We present an effective method to train multilingual IR systems when only English IR training data and some parallel corpora between English and other languages are available. 
We leverage parallel and non-parallel corpora to improve the pretrained multilingual language models’ cross-lingual transfer ability. 
%We consider both scenarios when we have parallel corpora between all languages with English, and when we do not have any parallel corpora for some low-resource languages. 
%We propose to train the dual-encoder dense passage retriever jointly with the semantic contrastive loss and the language contrastive loss. 
We design a \emph{semantic contrastive loss} to align representations of parallel sentences that share the same semantics in different languages, and a new \emph{language contrastive loss} to leverage parallel sentence pairs to remove language-specific information in sentence representations from non-parallel corpora. 
When trained on English IR data with these losses and evaluated zero-shot on non-English data, our model demonstrates significant improvement to prior work on retrieval performance, while it requires much less computational effort. 
We also demonstrate the value of our model for a practical setting when a parallel corpus is only available for a few languages, but a lack of parallel corpora resources persists for many other low-resource languages. 
Our model can work well even with a small number of parallel sentences,
%We also show that language contrastive loss can not only benefit the languages which do not have any parallel corpora, but also help the languages with parallel corpora, even the semantic contrastive loss has already been applied to them. 
and be used as an add-on module to any backbones and other tasks.

\let\thefootnote\relax\footnotetext{\noindent $^1$ Work done during an internship at AWS AI Labs. %Work done during Xiyang's internship at AWS AI.
}
\let\thefootnote\relax\footnotetext{\noindent * These authors contributed equally to this work.}
\let\thefootnote\relax\footnotetext{\noindent Code \href{https://github.com/xiyanghu/multilingualIR}{https://github.com/xiyanghu/multilingualIR}.}

\end{abstract}

\section{Introduction}

Information retrieval (IR) is an important natural language processing task that helps users efficiently gather information from a large corpus (some representative downstream tasks include question answering, summarization, search, recommendation, etc.), but developing effective IR systems for all languages is challenging due to the cost of, and therefore lack of, annotated training data in many languages. While this problem is not unique to IR research \cite{joshi-etal-2020-state}, constructing IR data is often more costly due to the need to either translate a large text corpus or gather relevancy annotations, or both, which makes it difficult to generalize IR models to lower-resource languages.

One solution to this is to leverage the pretrained multilingual language models to encode queries and corpora for multilingual IR tasks \cite{zhang-etal-2021-mr, sun-duh-2020-clirmatrix}.
One series of work on multilingual representation learning is based on training a masked language model, some with the next sentence prediction task, on monolingual corpora of many languages, such as mBERT and XLM-R \cite{conneau-etal-2020-unsupervised}. They generally do not explicitly learn the alignment across different languages and do not perform effectively in empirical IR experiments. Other works directly leverage multilingual parallel corpora or translation pairs to explicitly align the sentences in two languages, such as InfoXLM \cite{chi-etal-2021-infoxlm} and LaBSE \cite{feng-etal-2022-language}. %These works significantly improve the cross-language alignment, but are computationally heavy.
%\todo{add a sentence about the effect of these works and their potential drawbacks?}

\begin{figure}[t]
    \centering
    \includegraphics[width=0.47\textwidth]{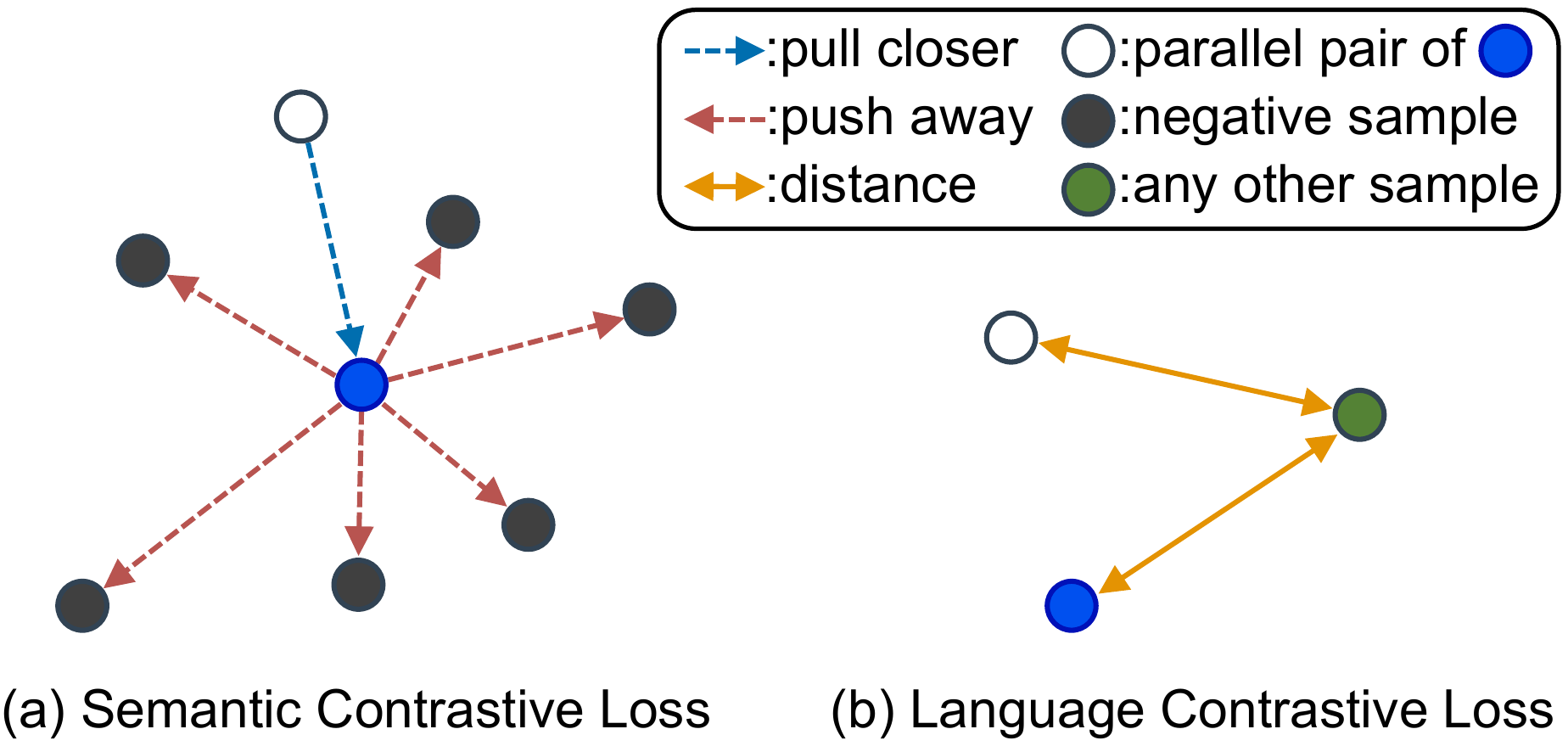}
    \vspace{-0.1in}
    \caption{
    (a) The \textit{semantic contrastive loss} encourages the embeddings of parallel pairs, i.e. sentences that have the same semantics but from different languages, to be close to each other, and away from the rest negative samples --- sentences with different semantics. (b) The \textit{language contrastive loss} incorporates the non-parallel corpora in addition to the parallel ones. It encourages the distances from a sentence representation, which can be a sample from both the parallel corpora and the non-parallel corpora, to the two embeddings of a paralleled pair to be the same.
%\todo{rephrase the intuition in the edited abstract here to answer the question ``why are we doing this''?}
    }
    \label{fig:diagram}
    \vskip -5mm
\end{figure}

In this work, we propose to use the \textit{semantic contrastive loss} and the \textit{language contrastive loss} to jointly train with the information retrieval objective, for learning cross-lingual representations that encourage efficient lingual transfer ability on retrieval tasks. Our semantic contrastive loss aims to align the embeddings of sentences that have the same semantics.
It is similar to the regular InfoNCE \cite{oord2018representation} loss, which forces the representations of parallel sentence pairs in two languages to be close to each other, and away from other negative samples. Our language contrastive loss aims to leverage the non-parallel corpora for languages without any parallel data, which are ignored by the semantic contrastive loss. 
It addresses the practical scenario wherein parallel corpora are easily accessible for a few languages, but the lack of such resources persists for many low-resource languages.
The language contrastive loss encourages the distances from a sentence representation to the two embeddings of a paralleled pair to be the same. Figure~\ref{fig:diagram} illustrates how the two losses improve language alignment. In experiments, we evaluate the zero-shot cross-lingual transfer ability of our model on monolingual information retrieval tasks for 10 different languages. Experimental results show that our proposed method obtains significant gains, and it can be used as an add-on module to any backbones. We also demonstrate that our method is much more computationally efficient than prior work. Our method works well with only a small number of parallel sentence pairs and works well on languages without any parallel corpora.

\begin{figure*}[t]
    \centering
    \includegraphics[width=0.95\textwidth]{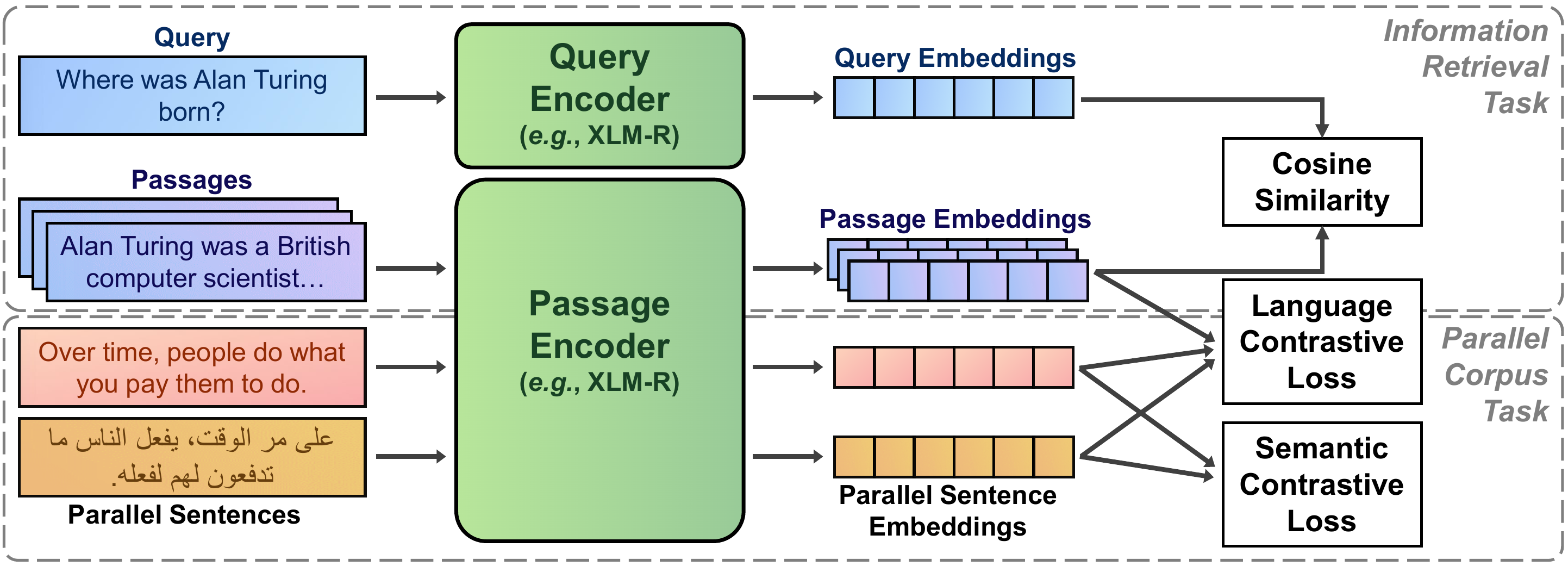}%{fig/model_framework.pdf}
    \vspace{-0.07in}
    \caption{
    Our model framework contains two parts: the main task (IR), and the parallel corpora task. For the main task part, we use a dual-encoder dense passage retrieval module for information retrieval. For the parallel corpora task part, we adopt the \textit{semantic contrastive loss} to improve cross-lingual domain adaptation with parallel corpora. We also use the \textit{language contrastive loss} by leveraging parallel corpora and non-parallel corpora altogether.
    }
    \label{fig:model}
\end{figure*}

\section{Background: Multilingual DPR}
\label{sec:mdpr}

 %\xccedit{}{Figure~\ref{fig:model} shows our model framework. There are two parts in our model: one is for our main retrieval task, and another is to leverage parallel and non-parallel corpora to improve cross-lingual adaptation.}

Dense Passage Retriever (DPR) \cite{karpukhin-etal-2020-dense} uses a dual-encoder structure to encode the queries and passages separately for information retrieval. To generalize to multilingual scenarios, we replace DPR's original BERT encoders with a multilingual language model XLM-R \cite{conneau-etal-2020-unsupervised} to transfer English training knowledge to other languages.

Concretely, given a batch of $N$ query-passage pairs ($\boldsymbol{p}_i$, $\boldsymbol{q}_i$), we consider all other passages $\boldsymbol{p}_j, j\neq i$ in the batch irrelevant (negative) passages, and optimize the retrieval loss function as the negative log-likelihood of the gold passage:
\begin{equation}
\small
\begin{aligned}
\mathcal{L}_{\text{IR}} &= -\frac{1}{N} \sum_{i=1}^N\\
& \log \frac{\exp \left(\operatorname{sim}\left(\boldsymbol{q}_{i}, \boldsymbol{p}_i\right) \right)}{\exp \left(\operatorname{sim}\left(\boldsymbol{q}_{i}, \boldsymbol{p}_i\right) \right) + \sum_{j=1, j\neq i}^N\exp \left(\operatorname{sim}\left(\boldsymbol{q}_{i}, \boldsymbol{p}_{j}\right) \right)}
\end{aligned}
\end{equation}
where the similarity of two vectors is defined as $\operatorname{sim}(\boldsymbol{u}, \boldsymbol{v})=\frac{\boldsymbol{u}^{\top} \boldsymbol{v}}{\|\boldsymbol{u}\|\|\boldsymbol{v}\|}
$. 

\section{Contrastive Learning for Cross-Lingual Generalization}
The multilingual dense passage retriever only uses English corpora for training. To improve the model's generalization ability to other languages, we leverage two contrastive losses, \textit{semantic contrastive loss} and \textit{language contrastive loss}. Figure~\ref{fig:model} shows our model framework.

Specifically, the \textit{semantic contrastive loss} \cite{chen2020simple} pushes the embedding vectors of a pair of parallel sentences close to each other, and at the same time away from other in-batch samples that have different semantics. The \textit{language contrastive loss} focuses on the scenario when there is no parallel corpora for some languages, which encourages the distance from a sentence embedding to paralleled embedding pairs to be the same.

% In addition to the contrastive loss above, we also consider the scenario when there is no parallel corpora for some languages. We propose the \textit{language contrastive loss} to train the model using the parallel corpora and non-parallel corpora altogether. The intuition is that we encourage the distance from a sentence embedding to paralleled embedding pairs to be the same.
%we also add an additional part to further remove the remaining language identity information in the embeddings. The intuition is that if we cannot classify which languages the embeddings are, then the domain/lingual shift problem are well-handled. Specifically, we follow \citet{ganin2015unsupervised} to use a \textbf{gradient reversal layer (GRL)} along with a classifier head to improve the domain adaptation. The gradient reversal layer works during the backpropagation by multiplying the gradients by a negative number (we use -1.0 in our model). During the training, we minimize the classification cross-entropy loss $\mathcal{L}_{langCLF}$ so that the classifier head would be optimized to be a strong classifier, while the upstream encoder will try to learn embeddings that are not helpful for language classification.

\subsection{Semantic Contrastive Loss}

%\todo{add intuition for why we're doing this, not just a textual explanation for what we are doing with the loss functions}
To learn a language-agnostic IR model, we wish to encode the sentences with the same semantics but from different languages to have the same embeddings. For each parallel corpora batch, we do not limit our sample to just one specific language pair. We randomly sample different language pairs for a batch. For example, a sampled batch could contain multiple language pairs of En-Ar, En-Ru, En-Zh, etc. This strategy can increase the difficulty of our contrastive learning and make the training more stable.

Concretely, we randomly sample a mini-batch of $2N$ data points ($N$ here does not have to be the same value as the $N$ in Section~\ref{sec:mdpr}). The batch contains $N$ pairs of parallel sentences from multiple different languages. Given a positive pair $\boldsymbol{z}_i$ and $\boldsymbol{z}_j$, the embedding vectors of a pair of parallel sentences $(i,j)$ from two languages, the rest $2(N-1)$ samples are used as negative samples. The semantic contrastive loss for a batch is:
\vskip -5mm
\begin{equation}
\small
\begin{aligned}
    \mathcal{L}_{\text{semaCL}} = &-\frac{1}{2N} \sum_{(i,j)}  \\
     {\big[} & \log \frac{\exp \left(\operatorname{sim}\left(\boldsymbol{z}_{i}, \boldsymbol{z}_{j}\right) / \tau\right)}{\sum_{k=1, k\ne i}^{2 N} \exp \left(\operatorname{sim}\left(\boldsymbol{z}_{i}, \boldsymbol{z}_{k}\right) / \tau\right)} + \\
    & \log \frac{\exp \left(\operatorname{sim}\left(\boldsymbol{z}_{j}, \boldsymbol{z}_{i}\right) / \tau\right)}{\sum_{k=1, k\ne j}^{2 N} \exp \left(\operatorname{sim}\left(\boldsymbol{z}_{j}, \boldsymbol{z}_{k}\right) / \tau\right)} {\big]}
\end{aligned}
\end{equation}

\noindent where $\tau$ is a temperature hyperparameter.

\subsection{Language Contrastive Loss}

%\todo{add intuition for why we're doing this, not just a textual explanation for what we are doing with the loss functions}

%The GRL+Classifier module could bring in several problems. First, the use of the classification head would introduce additional parameters. Second, the design of the classifier's structure might affect the performance. Third, such a GRL+Classifier is similar to adversarial training, which could require a different learning rate and parameter updating frequency from the main model. Therefore, we propose another way to encourage the encoder to remove language identity information in the embeddings.
% The problem with the semantic contrastive loss is that it can only be applied to languages with parallel datasets. A more prevalent 
When training multilingual IR systems, we might not always have parallel corpora for all languages of interest. In a realistic scenario, we have easy access to a few high-resource languages' parallel corpora, but no such availability for many low-resource languages. We propose a language contrastive loss to generalize the model's ability to the languages which do not have any parallel corpora. For a batch $B$ consisting of both parallel corpora $\mathbb{P}$ and non-parallel corpora $\mathbb{Q}$, we denote $\boldsymbol{z}_i$ and $\boldsymbol{z}_j$ as the embeddings of a pair of parallel sentences $(i,j)$ from two languages. We wish the cosine similarity from any other sentence embedding $\boldsymbol{z}_k$ to the two embeddings of a parallel pair to be the same. Therefore, we minimize the following loss.
\vskip -5mm
\begin{equation}
\small
\begin{aligned}
    \mathcal{L}_{\text{langCL}} = &-\frac{1}{N(N-2)} \sum_{(i,j) \in \mathbb{P}} \sum_{k \in (\mathbb{P} \cup \mathbb{Q}) \setminus \{i, j\}}  \\
    {\big[} &  \log \frac{\exp(\operatorname{sim}\left(\boldsymbol{z}_{i}, \boldsymbol{z}_{k}\right))}{\exp(\operatorname{sim}\left(\boldsymbol{z}_{i}, \boldsymbol{z}_{k}\right)) + \exp(\operatorname{sim}\left(\boldsymbol{z}_{j}, \boldsymbol{z}_{k}\right))} + \\
    & \log \frac{\exp(\operatorname{sim}\left(\boldsymbol{z}_{j}, \boldsymbol{z}_{k}\right))}{\exp(\operatorname{sim}\left(\boldsymbol{z}_{i}, \boldsymbol{z}_{k}\right)) + \exp(\operatorname{sim}\left(\boldsymbol{z}_{j}, \boldsymbol{z}_{k}\right))} \big]
    \label{eq:lang}
\end{aligned}
\end{equation}

\noindent The optimum can be reached when $\operatorname{sim}\left(\boldsymbol{z}_{i}, \boldsymbol{z}_{k}\right) = \operatorname{sim}\left(\boldsymbol{z}_{j}, \boldsymbol{z}_{k}\right)$ for all $i,j,k$. 
% The similarity of two vectors is defined as $\operatorname{sim}(\boldsymbol{u}, \boldsymbol{v})=\frac{\boldsymbol{u}^{\top} \boldsymbol{v}}{\|\boldsymbol{u}\|\|\boldsymbol{v}\|}$.
Note that the parallel corpus involved is not the target language’s parallel corpus. For example, in Formula~\ref{eq:lang}, $i$ and $j$ are two languages that are parallel with each other, and $k$ is a third language (target language) that does not have any parallel corpus with other languages.

\subsection{Semantic vs Language Contrastive Losses}
While both the semantic contrastive loss and language contrastive loss can serve to align the representations of parallel sentences and remove language bias, they achieve this goal differently, one via contrasting against in-batch negative samples, the other using in-batch parallel examples to constrain the target language embeddings.
Moreover, a key property of the language contrastive loss is that as long as there is some parallel corpus, we can use this loss function to remove the language bias from representations of sentences where no parallel data exists, which makes it more broadly applicable.
% Although the language contra stive loss has the same optimum as the semantic contrastive loss when only the parallel corpus is used, the optimization process is different, which could help the objective function look for the optimum points during the training. Moreover, the semantic contrastive loss only works on languages that have parallel corpora, while the language contrastive loss also involves the languages which do not have parallel corpora. 

\section{Training}

The two contrastive losses are applied to the passage encoder only. Experiments show that applying them to both the passage encoder and the query encoder would result in unstable optimization, where we see weird jumps in the training loss curves.

The joint loss with the information retrieval loss, the semantic contrastive loss, and the language contrastive loss is
\begin{equation}
\small
    \mathcal{L} = \mathcal{L}_{\text{IR}} + w_s \mathcal{L}_{\text{semaCL}} + w_l \mathcal{L}_{\text{langCL}},
    \label{eq:loss}
\end{equation}
where $w_s$ and $w_l$ are hyperparameters for the semantic contrastive loss and the language contrastive loss weights which need to be tuned adaptively in different tasks.

We train our model using 8 Nvidia Tesla V100 32GB GPUs. We use a batch size of 48. We use the AdamW optimizer with $\beta_1=0.9, \beta_2=0.999$ and a learning rate of $10^{-5}$. For the three losses $\mathcal{L}_{\text{IR}}, \mathcal{L}_{\text{semaCL}}, \mathcal{L}_{\text{langCL}}$, we sequentially calculate the loss and the gradients. We use $w_s=0.01$ and $w_l=0.001$. The hyperparameters are determined through a simple grid search.
%\todo{Add hyperparams settings here like batch size, learning rate, etc. Also, introduce training strategy here. There are three losses we train batch by batch. If necessary, we could have a algorithm to outline the process.}

% We denote $\boldsymbol{v}_{i1}$ and $\boldsymbol{v}_{i2}$ as the \textit{language} vectors of a pair of parallel sentences from language 1 and language 2.
% We wish the two language vectors of two sentences from the same language to be close to each other. Given a positive pair $\boldsymbol{v}_{il}$ and $\boldsymbol{v}_{jl}$ from the same language $l, l\in\{1,2\}$, all the rest $N$ samples of another language are used as negative samples.
% The contrastive loss of a positive pair $(\boldsymbol{v}_{i1}, \boldsymbol{v}_{j1})$ is:

% \begin{equation}
% \ell_{i1, j1}^{\text{lang}}=-\log \frac{\exp \left(\operatorname{sim}\left(\boldsymbol{v}_{i1}, \boldsymbol{v}_{j1}\right) / \tau\right)}{\exp \left(\operatorname{sim}\left(\boldsymbol{v}_{i1}, \boldsymbol{v}_{j1}\right) / \tau\right) + \sum_{k=1}^{N} \exp \left(\operatorname{sim}\left(\boldsymbol{v}_{i1}, \boldsymbol{v}_{k2}\right) / \tau\right)}
% %\label{eq:info-nce}
% \end{equation}

\section{Experiments}

\subsection{Datasets}
Our IR experiments involve two types of datasets: IR datasets and the parallel corpora.
\subsubsection{Information Retrieval}
In our experiments, we only use English information retrieval corpora (\textbf{Natural Questions}), and we evaluate the model's zero-shot transfer ability on other target languages (\textbf{Mr.TyDi}).
\begin{itemize}
    \item \textbf{Natural Questions} \cite{kwiatkowski-etal-2019-natural} is an English QA dataset. Following \citet{zhang-etal-2021-mr}, we use NQ dataset to train IR.
    \item \textbf{Mr.TyDi} is a multilingual dataset for monolingual retrieval \cite{zhang-etal-2021-mr}, which is constructed from a question answering dataset TyDi \cite{clark-etal-2020-tydi}. It contains eleven typologically diverse languages, i.e., Arabic (Ar), Bengali (Bn), English (En), Finnish (Fi), Indonesian (Id), Japanese (Ja), Korean (Ko), Russian (Ru), Swahili (Sw), Telugu (Te), Thai (Th). We do not use Mr.TyDi for IR training.
\end{itemize}
\subsubsection{Parallel Corpora} \textbf{WikiMatrix} parallel corpora contains extracted parallel sentences from the Wikipedia articles in 85 different languages \cite{schwenk-etal-2021-wikimatrix}. For those languages involved in the Mr.Tydi dataset, the number of parallel pairs between them and the English of the WikiMatrix dataset ranges from 51,000 and 1,019,000. During training, we sample the same number of parallel pairs (50K) for them.

\subsection{Baseline Models}

We apply our contrastive loss functions on three multilingual pretrained language models:
\begin{itemize}
\setlength\itemsep{.2em}
    \item \textbf{XLM-R} \cite{conneau-etal-2020-unsupervised} is a pre-trained transformer-based multilingual language model. It is trained on a corpus from 100 languages only with the Masked language Model (MLM) objective in a Roberta way.
    \item \textbf{InfoXLM} \cite{chi-etal-2021-infoxlm} uses 42GB parallel corpora to pre-train XLM-R by maximizing mutual information between multilingual-multi-granularity texts.
    \item \textbf{LaBSE} \cite{feng-etal-2022-language} pre-trains BERT with Masked Language Model and Translation Language Model on the monolingual data and bilingual translation pairs. They train the model by 1.8M steps using a batch size of 8192.
\end{itemize}

In Table~\ref{tab:models}, we compare the computational efforts needed by each model to improve the language transfer ability. Both InfoXLM and LaBSE require a large-scale pre-training which needs a larger batch size and a larger number of training steps than ours. Our model only requires "co-training" on the parallel corpora along with the main task. In Table~\ref{tab:models}, we list our model's training steps on the information retrieval task. This comparison indicates that  for the retrieval task, our model does not need the costly pre-training as InfoXLM and LaBSE.

\begin{table}[t]
\centering
\small
\begin{tabular}{lrrr}
\toprule
                                                                      & InfoXLM                                                             & LaBSE                                                           & Our Model                                                           \\ \midrule
\textbf{Batch Size}                                                   & 2,048                                                                & 8,192                                                            & 48                                                                  \\
\textbf{Training Steps}     & 200K                                                                & 1.8M                                                            & 24.54K                                                              \\
\textbf{Training Compute} & 347x & 12,518x & 1x \\ 
%\textbf{FLOPs}                                                   & 256$\times$15 TeraFLOPS                                                                & 512$\times$126 PetaFLOPS                                                            & 8$\times$15 TeraFLOPS                                                                  \\
\bottomrule
\end{tabular}
\vskip -1.5mm
\caption{A comparison of our model and baseline models' pre-training for lingual adaptation. Ours actually uses a "co-training" mode rather than "pre-training", so our training steps are the same as the main task. %\todo{what does this mean in FLOPs? Give at least a ballpark estimate}
}
\label{tab:models}
\vskip -3mm
\end{table}

\begin{table*}[t]
\begin{subtable}[h]{\textwidth}
\centering
\scriptsize
\begin{tabular}{lGllllllllllll}
\toprule
\textbf{Model}                                                                                             & \textbf{Metric} & \textbf{Ar}                   & \textbf{Bn}                   & \textbf{En}              & \textbf{Fi}                   & \textbf{Id}                   & \textbf{Ja}                   & \textbf{Ko}                            & \textbf{Ru}                            & \textbf{Sw}                   & \textbf{Te}                   & \textbf{Th}                            & \textbf{Avg} \\ \midrule
\textbf{XLM-R} & MRR         & 0.335                         & 0.345                         & 0.275                    & 0.302                         & 0.368                         & 0.274                         & 0.275                                  & 0.287                                  & 0.231                         & 0.298                         & 0.403                                  & 0.308        \\
\textbf{~~~ + semaCL} & MRR         & 0.399                         & \textbf{0.465}                         & 0.332                    & \textbf{0.355}                & \textbf{0.445}                & \textbf{0.360}                & \textbf{0.338}                         & \textbf{0.345}                                  & \textbf{0.281}                         & 0.550                & \textbf{0.482}                                  & \textbf{0.396}        \\
\textbf{~~~ + langCL} & MRR         & 0.402 &	0.437 &	\textbf{0.338} &	0.335 &	0.425 &	0.339 &	0.320 &	0.329 &	0.265 &	\textbf{0.600}	 & 0.453 &	0.386   \\

\textbf{~~~ + semaCL  + langCL} & MRR         & \textbf{0.404}                & \textbf{0.465}                         & \textbf{0.338}                    & 0.346                         & 0.430                         & 0.333                         & 0.320                                  & 0.341                                  & 0.266                         & 0.516                         & 0.477                                  & 0.385   \\ \bottomrule
\end{tabular}
\vskip -1.5mm
%\caption{MRR@100}
\label{tab:IR-MRR}
\end{subtable}
\vskip -1mm
\caption{MRR@100 on the monolingual information retrieval task of Mr.TyDi dataset.}
\label{tab:IR}
\end{table*}

\begin{table*}[t]
\begin{subtable}[h]{\textwidth}
\centering
\scriptsize
\begin{tabular}{lGllllllllllll}
\toprule
\textbf{Model}                                                                                             & \textbf{Metric} & \textbf{Ar}                   & \textbf{Bn}                   & \textbf{En}              & \textbf{Fi}                   & \textbf{Id}                   & \textbf{Ja}                   & \textbf{Ko}                            & \textbf{Ru}                            & \textbf{Sw}                   & \textbf{Te}                   & \textbf{Th}                            & \textbf{Avg} \\ 
\midrule
\multicolumn{14}{l}{\textit{Results reported by \citet{wu2022unsupervised}}} \\ 
\textbf{XLM-R}    & MRR             & 0.365       & 0.374       & 0.275       & 0.318       & 0.395       & 0.299       & 0.304       & 0.306       & 0.274       & 0.346       & 0.401       & 0.333        \\
\textbf{InfoXLM} & MRR             & 0.373       & 0.354       & 0.325       & 0.300         & 0.380        & 0.310        & 0.299       & 0.313       & 0.351       & 0.311       & 0.400         & 0.338        \\
\textbf{LABSE}   & MRR             & 0.372       & 0.504       & 0.314       & 0.309       & 0.376       & 0.271       & 0.309       & 0.325       & 0.394       & 0.465       & 0.374       & 0.365        \\
\textbf{CCP}   & MRR             & 0.426 & 0.457 & 0.359 & 0.372 & 0.462 & 0.377 & 0.346 & 0.360 & 0.392 & 0.470 & 0.489 & 0.410        \\
\midrule
\multicolumn{14}{l}{\textit{Results reported by \citet{zhang-etal-2021-mr}}} \\ 
\textbf{BM25 (default)}    & MRR             & 0.368  &  0.418  &  0.140   &   0.284   &    0.376    &  
 0.211    &    0.285    &    0.313    &    0.389    &    0.343    &   0.401    &    0.321    \\
\textbf{BM25 (tuned)} & MRR             & 0.367    &    0.413    &    0.151    &    0.288    &    0.382     &    0.217    &     0.281    &     0.329    &     0.396    &     0.424    &     0.417     &    0.333   \\
\midrule
{\textit{Our implementation}}\\
\textbf{XLM-R} & MRR         & 0.335                         & 0.345                         & 0.275                    & 0.302                         & 0.368                         & 0.274                         & 0.275                                  & 0.287                                  & 0.231                         & 0.298                         & 0.403                                  & 0.308        \\
\textbf{~~~ + semaCL} & MRR         & \textbf{0.399}                         & {0.465}                         & \textbf{0.332}                    & \textbf{0.355}                & \textbf{0.445}                & \textbf{0.360}                & \textbf{0.338}                         & \textbf{0.345}                                  & {0.281}                         & \textbf{0.550}                & \textbf{0.482}                                  & \textbf{0.396}        \\ 
\textbf{InfoXLM} & MRR         & 0.371                         & 0.337                         & 0.284                         & 0.272                         & 0.343                         & 0.311                         & 0.271                                  & 0.298                                  & {0.338}                         & {0.306}                         & 0.385                                  & 0.320       \\
\textbf{~~~ + semaCL} & MRR         & {0.375}                         & {0.413}                         & {0.331}                   & {0.314}                         & {0.406}                         & {0.321}                         & {0.292}                                  & {0.318}                                  & 0.299                         & 0.304                         & {0.427}                                  & {0.345} \\
\textbf{LaBSE} & MRR         & 0.321                         & 0.419                         & 0.240                         & 0.283                         & 0.347                         & {0.224}                         & {0.290}                                  & 0.296                                  & \textbf{0.428}                & {0.387}                         & {0.322}                                  & 0.323        \\
\textbf{~~~ + semaCL} & MRR         & {0.333}                         & \textbf{0.485}                         & {0.300}                   & {0.313}                         & {0.395}                         & 0.216                         & 0.265                                  & {0.329}                                  & 0.374                & 0.330                         & 0.308                                  & {0.332}        \\ 
 \bottomrule
\end{tabular}
\vskip -1.5mm
%\caption{MRR@100}
% \label{tab:IR-addon-MRR}
\end{subtable}
\vskip -1mm
\caption{MRR@100 on the monolingual information retrieval task of Mr.TyDi dataset.}
\label{tab:IR-addon}
\end{table*}

\subsection{Information Retrieval - All languages have parallel corpora with English}
\label{sec:parallel}

For the information retrieval training, we follow the previous literature \cite{zhang-etal-2021-mr, wu2022unsupervised} to use an English QA dataset -- the Natural Questions dataset \cite{kwiatkowski-etal-2019-natural} for both training and validation.

We evaluate our model performance on the Mr.TyDi dataset \cite{zhang-etal-2021-mr} for monolingual query passage retrieval in eleven languages. We follow \citet{zhang-etal-2021-mr} to use MRR@100 and Recall@100 as metrics.

In this section, we experimented with the setting when we have parallel corpora from English to all other target languages. We tested three different variants of our model using the XLM-R as the backbone: 
\begin{enumerate}
\setlength\itemsep{.1em}
    \item we only include the semantic contrastive loss for the parallel corpora: $\mathcal{L}_{\text{IR}} + w_s \mathcal{L}_{\text{semaCL}}$;
    \item we only include the language contrastive loss for the parallel corpora: $\mathcal{L}_{\text{IR}} + w_l \mathcal{L}_{\text{langCL}}$; %这里是w_l?
    %\item we use semantic contrastive loss and language classification loss with GRL: $\mathcal{L}_{\text{IR}} + w_s \mathcal{L}_{\text{semaCL}} + w_l \mathcal{L}_{langCLF}$;
    \item we use both the semantic contrastive loss and the language contrastive loss: $\mathcal{L}_{\text{IR}} + w_s \mathcal{L}_{\text{semaCL}} + w_l \mathcal{L}_{\text{langCL}}$.
\end{enumerate}

Table~\ref{tab:IR} shows the results of our model and the baseline XLM-R model. We also report the results of \citet{wu2022unsupervised}, which propose a model called contrastive context prediction (CCP) to learn multilingual representations by leveraging sentence-level contextual relations as self-supervision signals. For our analysis, we mainly focus on MRR, since MRR is more aligned with our retrieval loss function, which aims to rank relevant passages at higher orders. We also report Recall@100 in Table~\ref{tab:IR-Recall} in Appendix~\ref{sec:appendix}. We find that overall our model performs significantly better than the basic XLM-R. For our different model variants, we find that: (1) using only the semantic contrastive loss for the parallel corpora would achieve the best average performance; (2) using only the language contrastive loss for the parallel corpora also achieves a significant performance improvement, which is lower than but close to using only the semantic contrastive loss; (3) using both semantic contrastive loss and language contrastive loss would only contribute to a few languages like Ar, but does not improve the overall performance. Our assumption is that the semantic contrastive loss has already efficiently removed the language embedding shifts by leveraging the parallel pairs, so it is not helpful to use additional language contrastive loss when we have parallel corpora for all the languages. In Section~\ref{sec:nonparallel}, we experiment with a more practical scenario when we only have parallel corpora for some of the target languages but non-parallel corpora for the rest. And we find our language contrastive loss brings significant performance gains in that case.

We then further compare the performance of our best model --- XLM-R + semantic contrastive loss, with those of other strong baselines, i.e. InfoXLM and LaBSE. We also examine if the semantic contrastive loss can be used as an add-on module to InfoXLM and LaBSE to further boost their performance. Table~\ref{tab:IR-addon} shows the MRR@100 results of XLM-R, InfoXLM, LaBSE themselves --- all of them are trained with the IR loss, and the results trained jointly with the semantic contrastive loss. We find that our best model --- XLM-R with only semantic contrastive loss --- significantly outperforms these strong baselines. Note that both InfoXLM and LaBSE involve a large-scale pre-training to improve the lingual transfer ability, which is not required in our method. Our model only requires joint training with the contrastive loss, which needs much less computational effort as in Table~\ref{tab:models}. We also find that the semantic contrastive loss can be used as an add-on module to effectively boost the performance of InfoXLM and LaBSE. But such an add-on module's improvements on InfoXLM and LaBSE are not as large as that on XLM-R. We speculate that this phenomenon could be attributed to that InfoXLM and LaBSE have already been pre-trained on other datasets, which have some distribution shifts away from the WikiMatrix dataset we used for the semantic contrastive loss add-on module. We also report the Recall@100 results in Table~\ref{tab:IR-addon-Recall} of Appendix~\ref{sec:appendix}. In addition to the above results output by our own runs, we also list the results reported by \citet{wu2022unsupervised} in Table~\ref{tab:IR-addon} as a reference. The difference in the baseline model performances may be due to the randomness during model training. We also present the performance of the traditional BM25 method. The average MRR@100 of BM25 is significantly lower than that of our method.

\subsubsection{Effect of the Size of Parallel Dataset}

\begin{figure}[t]
    \centering
    % \vspace{-0.2in}
    \includegraphics[width=0.43\textwidth]{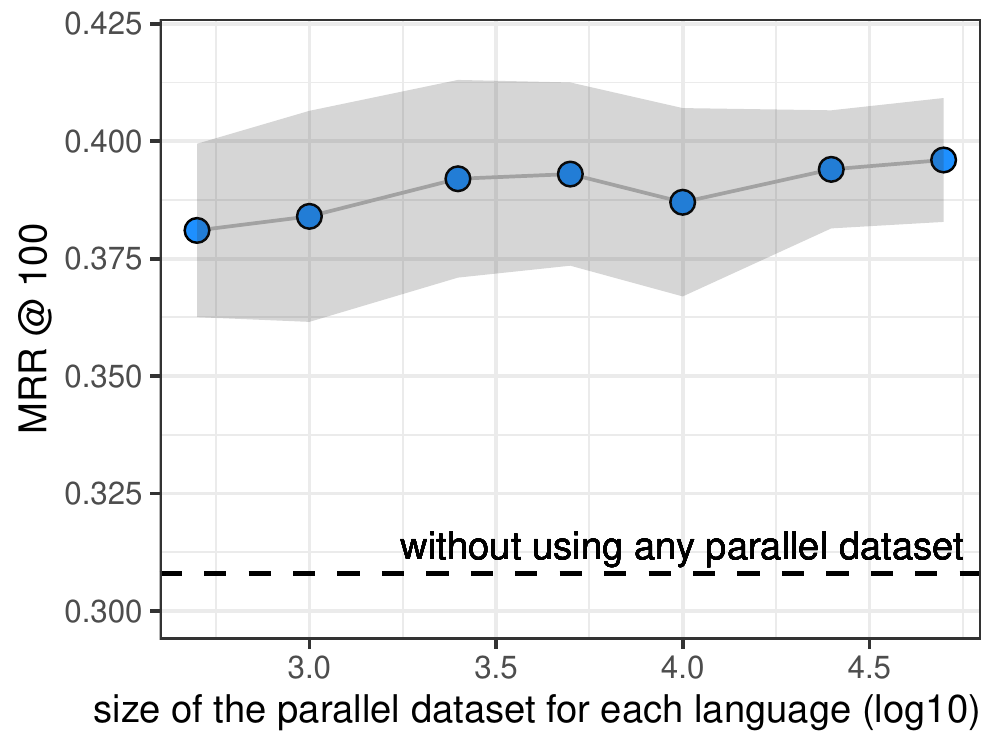}
    \vspace{-0.15in}
    \caption{
    The effect of the size of the parallel dataset for each language, with the 95\% CI in shadow. 
    }
    \label{fig:pds_mrr}
\end{figure}

We further investigate the effect of the size of the parallel dataset on the multilingual retrieval performance. We train our model by varying the parallel dataset size using the XLM-R with only semantic contrastive loss. Figure~\ref{fig:pds_mrr} shows the results. We find that: (1) using parallel corpora can significantly boost the retrieval performance, compared with the dashed horizontal line when we do not have parallel corpora at all (the basic XLM-R); (2) even when we only have a small parallel corpus of 500 pairs for each language, we can already achieve a good performance MRR@100=0.38. When we gradually increase the parallel corpora to 50,000, the MRR@100 grows gradually to 0.396. But the increase is not very large. This suggests that our model framework can work well even with a small parallel corpora dataset. This makes our method promising for those low-resource languages which lack parallel corpora with English.

\subsubsection{Effect of Language Pair Connection}
\label{sec:connect}

In order to understand how different language pair connections affect performance, we conduct experiments using different language pairs on En, Fi, Ja, Ko. We experimented with six different settings:
\begin{enumerate}
    \setlength\itemsep{.1em}
    \item \textbf{Basic setting:} Train XLM-R without using any parallel corpora, which is the same as the first row in Table~\ref{tab:IR};
    \item \textbf{Setting 1:} Train XLM-R with parallel corpora between English and all other languages, i.e. En-Fi, En-Ja, En-Ko;
    \item \textbf{Setting 2:} Train XLM-R with parallel corpora between English and Korean, and between Korean and the rest languages, i.e. En-Ko, Ko-Fi, Ko-Ja;
    \item \textbf{Setting 3:} Train XLM-R with parallel corpora between English and Korean, and between Japanese to Finnish, i.e. En-Ko, Ja-Fi;
    \item \textbf{Setting 4:} Train XLM-R with parallel corpora between English and Korean, i.e. En-Ko;
    \item \textbf{Setting 5:} Train XLM-R with parallel corpora  between Japanese to Finnish, i.e. Ja-Fi.
\end{enumerate}

Table~\ref{tab:connect} shows the MRR@100 results. We find that all settings 1 to 5 significantly surpass the basic setting. This echoes our previous finding that it is helpful to leverage parallel corpora. Among settings 1 to 5, the differences are not large --- the minimum MRR of them is 0.325, and the maximum one is 0.343. This suggests that the connection among language pairs is not a deterministic factor for our method. We also report the Recall@100 in Table~\ref{tab:connect-Recall} of Appendix~\ref{sec:appendix}.

\begin{table}[t]
\begin{subtable}[h]{0.48\textwidth}
\centering
\scriptsize
\begin{tabular}{lGlllll}
\toprule
                       & \textbf{Metric} & \textbf{En} & \textbf{Fi} & \textbf{Ja} & \textbf{Ko} & \textbf{Avg} \\ \midrule
\textbf{Basic Setting} & MRR             & 0.275       & 0.302       & 0.274       & 0.275       & 0.281        \\
\textbf{Setting 1}     & MRR             & 0.330       & 0.340       & 0.325       & 0.331       & 0.331        \\
\textbf{Setting 2}     & MRR             & 0.334       & 0.344       & \textbf{0.354}       & 0.320       & 0.338        \\
\textbf{Setting 3}     & MRR             & 0.318       & 0.341       & 0.348       & \textbf{0.365}       & \textbf{0.343}        \\
\textbf{Setting 4}     & MRR             & \textbf{0.336}       & \textbf{0.354}       & 0.340       & 0.345       & \textbf{0.343}        \\
\textbf{Setting 5}     & MRR             & 0.323       & 0.352       & 0.317       & 0.311       & 0.325        \\ \bottomrule
\end{tabular}
\vskip -1.5mm
%\caption{MRR@100}
\label{tab:connect-MRR}
\end{subtable}
% \\
% \begin{subtable}[h]{0.48\textwidth}
% \centering
% \scriptsize
% \begin{tabular}{lGlllll}
% \toprule
%                        & \textbf{Metric} & \textbf{En} & \textbf{Fi} & \textbf{Ja} & \textbf{Ko} & \textbf{Avg} \\ \midrule
% \textbf{Basic Setting} & Recall          & 0.754       & 0.755       & 0.741       & 0.691       & 0.735        \\
% \textbf{Setting 1}     & Recall          & 0.776       & 0.770       & 0.777       & 0.706       & 0.757        \\
% \textbf{Setting 2}     & Recall          & \textbf{0.785}       & 0.778       & \textbf{0.781}       & 0.710       & \textbf{0.763}        \\
% \textbf{Setting 3}     & Recall          & 0.767       & 0.762       & 0.766       & \textbf{0.723}       & 0.754        \\
% \textbf{Setting 4}     & Recall          & 0.779       & \textbf{0.785}       & 0.764       & 0.722       & 0.762        \\
% \textbf{Setting 5}     & Recall          & 0.765       & 0.759       & 0.722       & 0.703       & 0.737        \\ \bottomrule
% \end{tabular}
% \vskip -1.5mm
% \caption{Recall@100}
% \label{tab:connect-Recall}
% \end{subtable}
\vskip -1mm
\caption{MRR@100 on different language pair connections.}
\label{tab:connect}
\vskip -4mm
\end{table}

% \usepackage{multirow}
\begin{table*}[!htbp]
\centering
\scriptsize

\begin{subtable}[h]{\textwidth}
\begin{threeparttable}
\begin{tabular}{lG|p{0.11in}p{0.11in}p{0.11in}p{0.11in}p{0.11in}p{0.11in}p{0.11in}l|p{0.11in}p{0.11in}p{0.11in}p{0.11in}l||l}
\toprule
                                                                               \textbf{Model}                          & \textbf{Metric} & \textbf{Ar} & \textbf{Bn} & \textbf{En} & \textbf{Fi} & \textbf{Id} & \textbf{Ja} & \textbf{Ko} & \textbf{Avg\tnote{$\parallel$}} & \textbf{Ru} & \textbf{Sw} & \textbf{Te} & \textbf{Th} & \textbf{Avg\tnote{$\nparallel$}} & \textbf{Avg} \\ \midrule
\textbf{XLM-R} & MRR         & 0.335                         & 0.345                         & 0.275                    & 0.302                         & 0.368                         & 0.274                         & 0.275      &0.307                            & 0.287                                  & 0.231                         & 0.298                         & 0.403     &0.305                             & 0.308        \\
\textbf{~~~ + semaCL}                                                                 & MRR             & 0.389       & 0.427       & 0.310        & 0.333       & 0.404       & 0.323       & 0.316       & 0.365                                                                & 0.319       & 0.252       & 0.423       & 0.448       & 0.360                                                                  & 0.358                                                        \\ 
\textbf{~~~ + langCL (WikiMatrix)}    & MRR             & 0.385       & 0.394        & 0.314       & 0.340       & 0.411       & 0.319       & 0.299       & 0.351                                                                & 0.316       & 0.227       & 0.474       & 0.430       & 0.361                                                                 & 0.355                                                        \\
\textbf{~~~ + langCL (Mr.TyDi)}    & MRR             & 0.347       & 0.378        & 0.287       & 0.306       & 0.371       & 0.292       & 0.277       & 0.322                                                                & 0.294       & 0.221       & 0.248       & 0.395       & 0.289                                                                 & 0.311                                                        \\
\textbf{~~~ + semaCL + langCL (WikiMatrix)} & MRR             & 0.396       & 0.421       & \textbf{0.342}       & 0.357       & \textbf{0.438}       & \textbf{0.357}       & \textbf{0.336}       & 0.384                                                                & \textbf{0.350}        & 0.274       & 0.420        & \textbf{0.476}       & 0.380                                                                  & 0.378                                                        \\
                                                                                                        %  &           &        &        &        &        &        &        &        &        &        &        &        &        &                                                                  &                                                        \\  \midrule
\textbf{~~~ + semaCL + langCL (Mr.TyDi)}    & MRR             & \textbf{0.408}       & \textbf{0.470}        & 0.336       & \textbf{0.362}       & \textbf{0.438}       & 0.339       & 0.332       & \textbf{0.391}                                                                & 0.347       & \textbf{0.291}       & \textbf{0.449}       & 0.471       & \textbf{0.389}                                                                 & \textbf{0.385}                                                        \\
                                                                                                        %  &           &        &        &        &        &        &        &        &        &        &        &        &        &                                                                  &                                                        \\  
                                                   \bottomrule
\end{tabular}
\begin{tablenotes}
    \item[] Note: Avg for languages with ($\parallel$) and without ($\nparallel$) parallel data.
  \end{tablenotes}
\end{threeparttable}
\vskip -1.5mm
%\caption{MRR@100}
\label{tab:IR-nonpara-MRR}
\end{subtable}

\vskip -1mm
\caption{Experiment results when Ru, Sw, Te, Th do NOT have parallel data (MRR@100).}
\label{tab:IR-nonparallel}
\vskip -2mm
\end{table*}

\subsection{Information Retrieval - Some languages do not have parallel data}
\label{sec:nonparallel}

In this section, we investigate the scenario when we have parallel corpora only for some of the target languages, but not for the rest languages. This scenario emphasizes a realistic constraint that we lack parallel corpora for many low-resource languages. To test it, we leave Ru, Sw, Te, Th as languages that do not have parallel corpora, and keep these parallel corpora for all other languages, i.e. Ar, Bn, Fi, Id, Ja, Ko. We experimented with three different settings:
\begin{enumerate}
\setlength\itemsep{.1em}
    \item \textbf{XLM-R + Semantic CL:} we only use the semantic contrastive loss on languages which have parallel corpora (Ar, Bn, Fi, Id, Ja, Ko): $\mathcal{L}_{\text{IR}} + w_s \mathcal{L}_{\text{semaCL}}$;
    \item \textbf{XLM-R + Semantic CL + Language CL (WikiMatrix):} we use the semantic contrastive loss on languages which have parallel corpora (Ar, Bn, Fi, Id, Ja, Ko), and the language contrastive loss on these parallel corpora (Ar, Bn, Fi, Id, Ja, Ko) along with the non-parallel WikiMatrix corpora: $\mathcal{L}_{\text{IR}} + w_s \mathcal{L}_{\text{semaCL}} + w_l \mathcal{L}_{\text{langCL}}$.
    \item \textbf{XLM-R + Semantic CL + Language CL (Mr.TyDi):} we use the semantic contrastive loss on languages which have parallel corpora (Ar, Bn, Fi, Id, Ja, Ko), and the language contrastive loss on these parallel corpora (Ar, Bn, Fi, Id, Ja, Ko) along with the non-parallel Mr.TyDi corpora: $\mathcal{L}_{\text{IR}} + w_s \mathcal{L}_{\text{semaCL}} + w_l \mathcal{L}_{\text{langCL}}$;
\end{enumerate}

Table~\ref{tab:IR-nonparallel} shows the MRR@100 results of our experiments. The language contrastive loss can effectively leverage the non-parallel corpora to improve the information retrieval performance. For the \textbf{XLM-R + Semantic CL + Language CL (Mr.TyDi)} setting, the language contrastive loss boosts the average MRR@100 from 0.358 to 0.385. We also calculate the average performance on the languages with parallel corpora (Ar, Bn, Fi, Id, Ja, Ko), and the languages without parallel corpora (Ru, Sw, Te, Th). The \textbf{Avg (withParallel)} column and the \textbf{Avg (noParallel)} column in Table~\ref{tab:IR-nonparallel} are their corresponding results. We find that the language contrastive loss can improve the performance on both types of languages. For languages with parallel corpora (Ar, Bn, Fi, Id, Ja, Ko), the MRR@100 increases from 0.365 to 0.391; for languages without parallel corpora (Ru, Sw, Te, Th), the MRR@100 increase from 0.360 to 0.389. This result suggests our model can be effectively deployed in situations when we have no parallel corpora for low-resource languages. Appendix~\ref{sec:appendix} Table~\ref{tab:IR-nonpara-Recall} reports the Recall@100 results.

Since using the Mr.TyDi corpora brings in the target domain information, we also examine the \textbf{XLM-R + Semantic CL + Language CL (WikiMatrix)} setting. This setting uses the WikiMatrix non-parallel corpora for Ru, Sw, Te, Th --- it does not introduce the target domain information, and reflects the clean gain from the language contrastive loss. We find that using the WikiMatrix non-parallel corpora achieves a little lower but close performance than the one using the Mr.TyDi corpora. This suggests that the introduction of the target domain information is very minor in improving IR performance.

\subsubsection{Effect of the Size of Non-Parallel Dataset}

\begin{figure}[t]
    \centering
    \vspace{-0.1in}
    \includegraphics[width=0.43\textwidth]{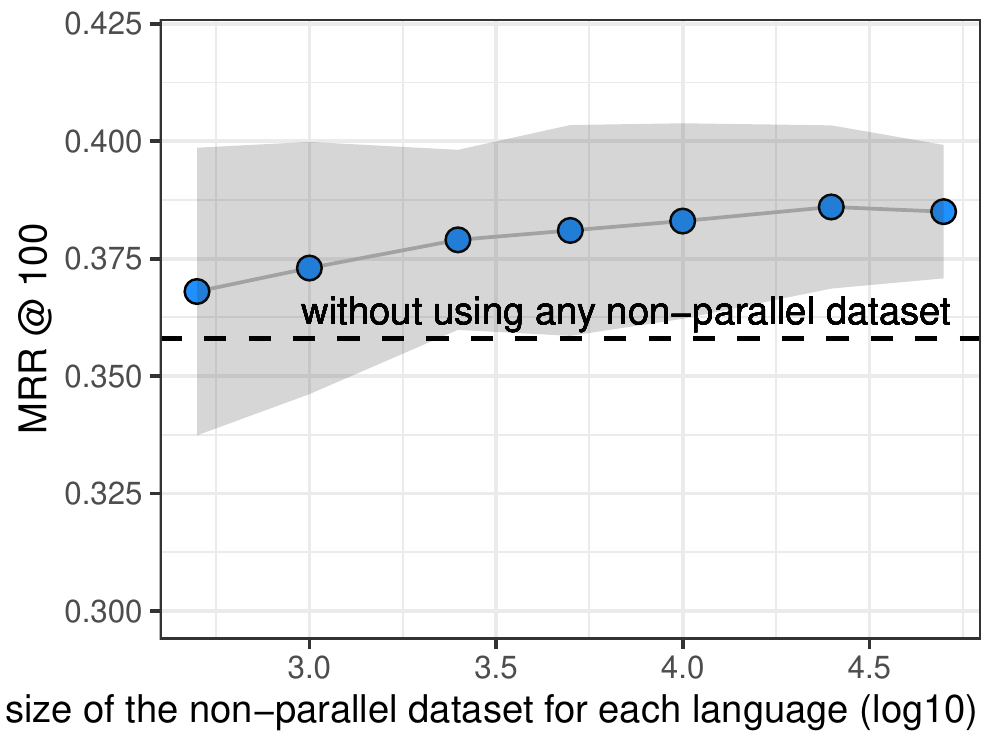}
    \vspace{-0.15in}
    \caption{
    The effect of the size of the non-parallel dataset for each language, with the 95\% CI in shadow. 
    }
    \label{fig:npds_mrr}
    \vskip -4mm
\end{figure}

We further investigate the effect of the size of the non-parallel dataset on the multilingual retrieval performance. We train our model by varying the non-parallel dataset size using XLM-R with both the semantic contrastive loss and the language contrastive loss. We keep the size of the parallel dataset fixed at 50,000. Figure~\ref{fig:npds_mrr} shows the results. The dashed horizontal line is the one using only parallel corpora, i.e. the first row in Table~\ref{tab:IR-nonparallel}. We find that: (1) using non-parallel corpora can significantly boost the retrieval performance, compared with the most left point when we do not use the non-parallel corpora at all; (2) when the non-parallel corpora dataset size increases from 0 to 10,000, the MRR@100 improves quickly; (3) when the non-parallel dataset size increases from 10,000 to 50,000, the MRR@100 has minor changes, but its variance decreases.

\subsection{BUCC: Bitext Retrieval}

The information retrieval task above is not common to see in multilingual NLP papers. A closely related task they often work on is the BUCC\footnote{\textbf{BUCC} (Building and Using Comparable Corpora) is a dataset for bitext mining --- identify sentence pairs that are translations in two different languages \cite{zweigenbaum2018overview}. The task aims to mine parallel sentences between English and four other languages, i.e., German, French, Russian, and Chinese. About 2–3\% of the sentences in the whole dataset are gold parallel.} task \cite{zweigenbaum2018overview}. The BUCC task has been tested in the LaBSE benchmark model we used in the previous section \cite{feng-etal-2022-language}, and in many other multilingual NLP works, such as \citealt[etc]{artetxe-schwenk-2019-margin, ijcai2019p746, schwenk-2018-filtering, reimers-gurevych-2020-making}. Therefore, following these works, we also investigate our model's performance on the BUCC task. 

For the BUCC bitext mining task, we follow previous work \cite{artetxe-schwenk-2019-margin, reimers-gurevych-2020-making} to first encode texts, and then use the equation below to calculate the score of two sentence embeddings $\boldsymbol{u}, \boldsymbol{v}$:
\begin{equation}
\scriptsize
\begin{aligned}
    \operatorname{score}(\boldsymbol{u}, \boldsymbol{v})=
    \frac{\operatorname{sim}(\boldsymbol{u}, \boldsymbol{v})}{\sum_{\boldsymbol{z} \in \operatorname{NN}_k(\boldsymbol{u})} \frac{\operatorname{sim}(\boldsymbol{u}, \boldsymbol{z})}{2k} + \sum_{\boldsymbol{z} \in \operatorname{NN}_k(\boldsymbol{v})} \frac{\operatorname{sim}(\boldsymbol{v}, \boldsymbol{z})}{2k}}
\end{aligned}
\end{equation}

\noindent where $\operatorname{NN}_k(\boldsymbol{u})$ denotes $\boldsymbol{u}$'s $k$ nearest neighbors in another language. The training set is used to find a threshold value of the score, for which pairs with scores above this threshold are predicted as parallel sentences. We use F1 to measure the model performance on BUCC.

Table~\ref{tab:bucc} shows F1 score of our model based on the XLM-R, InfoXLM, and LaBSE. We first examine the vanilla XLM-R, InfoXLM, and LaBSE as the text encoder. LaBSE and InfoXLM outperform XLM-R a lot due to their large-scale pre-training on improving lingual adaptation using parallel datasets. When we add our semantic contrastive loss to XLM-R, we get a large improvement across all four languages. We find that our model (XLM-R + Semantic CL) outperforms XLM-R, but underperforms InfoXLM and LaBSE. We attribute LaBSE's great performance to its much larger pre-training than ours, and LaBSE's training involves Translation Language Model \cite{conneau2019cross} with translation corpora. This is exactly the same type of corpora as BUCC's translation parallel pairs. When we add our semantic contrastive loss to InfoXLM, we obtain performance gain for all languages. The gain is smaller than that of XLM-R because InfoXLM has already been trained on parallel corpora. When we add our semantic contrastive loss module to LaBSE, we obtained a small increase in the average performance --- the performance on Zh has a significant increase.

\begin{table}
\centering
\scriptsize
\begin{tabular}{p{1.05in}p{0.16in}p{0.16in}p{0.16in}p{0.16in}p{0.16in}}
\toprule
\textbf{Model}                    & \multicolumn{1}{l}{\textbf{De}} & \multicolumn{1}{l}{\textbf{Fr}} & \multicolumn{1}{l}{\textbf{Ru}} & \multicolumn{1}{l}{\textbf{Zh}} & \multicolumn{1}{l}{\textbf{Avg}} \\ \midrule
\textbf{XLM-R}               & 17.90                           & 12.12                           & 21.84                           & 15.06            & 16.73              \\
\textbf{~~~ + semaCL} & 72.86                           & 69.44                           & 73.10                           & 66.27            & 70.42              \\ \midrule
\textbf{InfoXLM}             & 63.32                           & 54.36                           & 69.92                           & 66.19            & 63.45              \\
\textbf{~~~ + semaCL}             & 80.40                           & 75.48                           & 78.20                           & 75.04            & 77.28              \\ \midrule
\textbf{LaBSE} \cite{feng-etal-2022-language}               & 92.50                           & 88.70                           & 88.90                           & 88.90            & 89.75              \\
\textbf{LaBSE}               & \textbf{93.03}                           & \textbf{89.75}                           & 89.75                           & 85.93            & 89.62              \\
\textbf{~~~ + semaCL}               & 92.95                           & 89.50                           & \textbf{89.82}                           & \textbf{89.77}            & \textbf{90.51}              \\
\bottomrule
\end{tabular}
\vskip -2mm
\caption{F1 score on the BUCC task. Row \textbf{LaBSE} \cite{feng-etal-2022-language} is the results reported in \citet{feng-etal-2022-language}. All other rows are the output of our own implementation.}
\label{tab:bucc}
\vskip -3mm
\end{table}

One important insight we get from comparing Table~\ref{tab:bucc} and Table~\ref{tab:IR-addon} is that a model's better performance in NLP tasks like BUCC does not necessarily mean better performance in information retrieval. Most existing multilingual NLP papers only examine the BUCC bi-text retrieval task, and we highlight the inconsistency between models' performances on the two types of retrieval tasks.

\section{Related Work}
%\todo{Related work session in my opinion (xcc) should be organized in a better way that show the relationship of those content with our work and the relation of parts should be organized too. Like you can split IR into self-supervise / supervise models or monolingual / multilingual models, then describe the related work accordingly.}

Dense monolingual / multilingual information retrieval study recently attracts great attention, which mainly benefits from (1) supervised finetuning based on large pre-trained language models and (2) self-supervised contrastive learning. 

Dense Passage Retrieval \cite{karpukhin-etal-2020-dense} is the framework first proposed for monolingual \textbf{superivsed finetuning} on information retrieval. It uses a BERT-based dual-encoder structure to encode the query and the candidate passages into embeddings. Similar to monolingual IR, supervised finetuning can also be applied to  multilingual pretrained language models (LMs) for multilingual IR. Commonly uesd multilingual pretrained LMs include multilingual BERT (mBERT, \citealp{devlin-etal-2019-bert}) and XLM-R \cite{conneau-etal-2020-unsupervised}, both of which are trained on large corpora representing about 100 languages primarily with the masked language modeling task. These models do not use any explicit objective to improve the alignment between language sentences. Recent efforts in NLP field have provided easy access to parallel corpora, e.g. \citet{schwenk-etal-2021-wikimatrix}. Many  multilingual language models use additional parallel data to improve lingual transfer ability. InfoXLM \cite{chi-etal-2021-infoxlm} uses parallel corpora to pre-train  XLM-R by maximizing mutual information between multilingual multi-granularity texts. LaBSE \cite{feng-etal-2022-language} pre-trains BERT with Masked Language  Model and Translation Language Model on the monolingual data and bilingual translation pairs.

\textbf{Self-supervised contrastive learning} is another way used to improve the cross-lingual alignment. 
Contrastive learning maximizes the agreement between positive samples, and minimizes the similarity of positive and negative ones \citep{he2020momentum, chen2020simple, chen2020big, chen2020mocov2}. 
%In NLP area, researchers recently started to adapt contrastive learning to various tasks.
For language representation learning, \citet{clark2019electra} apply contrastive learning to train a discriminative model to learn language representations. 
%For text domain adaptation, \citet{li2020cross} adopt contrastive learning for cross-domain sentiment classification. 
For multilingual representation learning, contrastive learning has been used to improve cross-lingual transfer ability by using additional parallel data \cite{hu-etal-2021-explicit} or by leveraging other self-supervision signals \cite{wu2022unsupervised}.

\section{Conclusion}

In this paper, we present a model framework for multilingual information retrieval by improving lingual adaptation through contrastive learning. %Our semantic contrastive loss learns the language alignment based on parallel corpora. Our language contrastive loss further improves the lingual transfer ability by leveraging both the parallel corpora and the non-parallel corpora. 
Our experiments demonstrate the effectiveness of our methods in learning better cross-lingual representations for information retrieval tasks.
%Our method does not require time-consuming pre-training and is significantly computationally friendly than prior work. 
The two contrastive losses can be used as an add-on module to any backbones and many other tasks besides information retrieval. 
%Future work can further investigate the heterogeneous effects of our methods on different languages, and how different corpora language pairs' parallel corpora affect the cross-lingual transfer ability.

\section{Limitations}

In this work, we did not conduct a detailed analysis of how language-specific characteristics contribute to our model's cross-lingual generalization capabilities. Future work may address this question through extensive matrix experiments --- traverse the training on each possible language pair combination and evaluate on all languages.

\bibliography{anthology,custom}
\bibliographystyle{acl_natbib}

%\newpage

\appendix
\clearpage
\section{Appendix}
\label{sec:appendix}
Table~\ref{tab:IR-Recall} and Table~\ref{tab:IR-addon-Recall} show the Recall@100 results of experiments in Section~\ref{sec:parallel}.
Table~\ref{tab:connect-Recall} shows the Recall@100 results of experiments in Section~\ref{sec:connect}.
Table~\ref{tab:IR-nonpara-Recall} shows the Recall@100 results of experiments in Section~\ref{sec:nonparallel}.
%
%This is an appendix.
\begin{table*}[!h]
\begin{subtable}[h]{\textwidth}
\centering
\scriptsize
\begin{tabular}{lGllllllllllll}
\toprule
\textbf{Model}                                                                                             & \textbf{Metric} & \textbf{Ar}                   & \textbf{Bn}                   & \textbf{En}              & \textbf{Fi}                   & \textbf{Id}                   & \textbf{Ja}                   & \textbf{Ko}                            & \textbf{Ru}                            & \textbf{Sw}                   & \textbf{Te}                   & \textbf{Th}                            & \textbf{Avg} \\ \midrule
\textbf{XLM-R} & Recall      & 0.782                         & 0.797                         & 0.754                    & 0.755                         & 0.840                         & 0.741                         & 0.691                                  & 0.741                                  & 0.614                         & 0.820                         & 0.852                                  & 0.762        \\
\textbf{~~~ + semaCL} & Recall      & 0.799                         & 0.851                         & 0.777                   & 0.773                         & \textbf{0.867}                & 0.779                         & 0.730                                  & \textbf{0.763}                         & 0.597                         & 0.862                         & \textbf{0.886}                                  & 0.789        \\
\textbf{~~~ + langCL} & Recall      & 0.799 &	0.820 &	0.782 &	0.769 &	0.858 &	0.769 &	0.723 &	0.750 &	\textbf{0.629} &	\textbf{0.892} &	0.871 &	0.788   \\
\textbf{~~~ + semaCL + langCL} & Recall      & \textbf{0.806}                & \textbf{0.864}                         & \textbf{0.798}             & \textbf{0.784}                & 0.858                         & \textbf{0.780}                         & \textbf{0.736}                                  & 0.743                                  & 0.626                         & 0.867                         & 0.877                                  & \textbf{0.794}   \\ \bottomrule
\end{tabular}
\vskip -1.5mm
%\caption{Recall@100}
\end{subtable}
\vskip -1mm
\caption{Recall@100 on the monolingual information retrieval task of Mr.TyDi dataset.}
%\label{tab:IR}
\label{tab:IR-Recall}
\end{table*}

\begin{table*}[t]
\begin{subtable}[h]{\textwidth}
\centering
\scriptsize
\begin{tabular}{lGllllllllllll}
\toprule
\textbf{Model}                                                                                             & \textbf{Metric} & \textbf{Ar}                   & \textbf{Bn}                   & \textbf{En}              & \textbf{Fi}                   & \textbf{Id}                   & \textbf{Ja}                   & \textbf{Ko}                            & \textbf{Ru}                            & \textbf{Sw}                   & \textbf{Te}                   & \textbf{Th}                            & \textbf{Avg} \\ \midrule
\multicolumn{14}{l}{\textit{Results reported by \citet{wu2022unsupervised}}} \\ 
\textbf{XLM-R}    & Recall          & 0.813       & 0.842       & 0.776       & 0.782       & 0.886       & 0.785       & 0.727       & 0.774       & 0.633       & 0.875       & 0.882       & 0.798        \\
\textbf{InfoXLM} & Recall          & 0.806       & 0.860        & 0.804       & 0.749       & 0.869       & 0.788       & 0.717       & 0.767       & 0.724       & 0.867       & 0.874       & 0.802        \\
\textbf{LABSE}   & Recall          & 0.762       & 0.910        & 0.783       & 0.760        & 0.852       & 0.669       & 0.644       & 0.744       & 0.750        & 0.889       & 0.834       & 0.782        \\
\textbf{CCP}   & Recall          & 0.820 & 0.883 & 0.801 & 0.787 & 0.875 & 0.800 & 0.732 & 0.772 & 0.751 & 0.888 & 0.889 & 0.818        \\ \midrule
\multicolumn{14}{l}{\textit{Results reported by \citet{zhang-etal-2021-mr}}} \\ 
\textbf{BM25 (default)}    & MRR             & 0.793  &  0.869  &  0.537   &   0.719   &    0.843    &  
 0.645    &    0.619    &    0.648    &    0.764    &    0.758    &   0.853    &    0.732    \\
\textbf{BM25 (tuned)} & MRR             & 0.800    &    0.874    &    0.551    &    0.725    &    0.846     &    0.656    &     0.797    &     0.660    &     0.764    &     0.813    &     0.853     &    0.758   \\
\midrule
{\textit{Our implementation}}\\
\textbf{XLM-R} & Recall      & 0.782                         & 0.797                         & 0.754                    & 0.755                         & 0.840                         & 0.741                         & 0.691                                  & 0.741                                  & {0.614}                         & 0.820                         & 0.852                                  & 0.762        \\
\textbf{~~~ + semaCL} & Recall      & \textbf{0.799}                         & {0.851}                         & {0.777}                   & \textbf{0.773}                         & \textbf{0.867}                & {0.779}                         & \textbf{0.730}                                  & {0.763}                         & 0.597                         & {0.862}                         & \textbf{0.886}                                  & {0.789}       \\ %\midrule
\textbf{InfoXLM} & Recall      & 0.797                        & \textbf{0.900}                & 0.785                  & 0.725                         & {0.843}                         & 0.790                & {0.717}                                  & {0.753}                                  & {0.711}                         & \textbf{0.873}                         & {0.875}                                  & \textbf{0.797}    \\
\textbf{~~~ + semaCL} & Recall      & 0.790                         & 0.842                & \textbf{0.791}                   & {0.731}                         & 0.829                         & \textbf{0.805}                & 0.708                                  & {0.753}                                  & 0.646                         & 0.800                         & 0.866                                  & 0.778       \\ %\midrule
\textbf{LaBSE} & Recall      & 0.769                         & {0.887}                         & 0.760                    & {0.773}                         & {0.854}                         & {0.652}                         & {0.649}                                  & \textbf{0.764}                                  & \textbf{0.832}                & {0.862}                & {0.824}                                  & {0.784}        \\
\textbf{~~~ + semaCL} & Recall      & 0.725                         & 0.865                         & {0.762}                   & 0.770                         & 0.845                         & 0.575                         & 0.604                                  & 0.734                                  & 0.739                & 0.724                & 0.697                                  &   0.731   \\
\bottomrule
\end{tabular}
\vskip -1.5mm
%\caption{Recall@100}
\end{subtable}
\vskip -1mm
\caption{Recall@100 on the monolingual information retrieval task of Mr.TyDi dataset.}
%\label{tab:IR-addon}
\label{tab:IR-addon-Recall}
\end{table*}

\begin{table*}[!h]
\centering
\begin{subtable}[h]{0.48\textwidth}
\scriptsize
\begin{tabular}{lGlllll}
\toprule
                       & \textbf{Metric} & \textbf{En} & \textbf{Fi} & \textbf{Ja} & \textbf{Ko} & \textbf{Avg} \\ \midrule
\textbf{Basic Setting} & Recall          & 0.754       & 0.755       & 0.741       & 0.691       & 0.735        \\
\textbf{Setting 1}     & Recall          & 0.776       & 0.770       & 0.777       & 0.706       & 0.757        \\
\textbf{Setting 2}     & Recall          & \textbf{0.785}       & 0.778       & \textbf{0.781}       & 0.710       & \textbf{0.763}        \\
\textbf{Setting 3}     & Recall          & 0.767       & 0.762       & 0.766       & \textbf{0.723}       & 0.754        \\
\textbf{Setting 4}     & Recall          & 0.779       & \textbf{0.785}       & 0.764       & 0.722       & 0.762        \\
\textbf{Setting 5}     & Recall          & 0.765       & 0.759       & 0.722       & 0.703       & 0.737        \\ \bottomrule
\end{tabular}
\vskip -1.5mm
%\caption{Recall@100}
\end{subtable}
\vskip -1mm
\caption{Recall@100 on different language pair connections.}
%\label{tab:connect}
\label{tab:connect-Recall}
\end{table*}

\begin{table*}[!h]
\centering
\scriptsize
\begin{subtable}[h]{\textwidth}
\begin{threeparttable}
\begin{tabular}{lG|p{0.11in}p{0.11in}p{0.11in}p{0.11in}p{0.11in}p{0.11in}p{0.11in}l|p{0.11in}p{0.11in}p{0.11in}p{0.11in}l||l}
\toprule
                                                                               \textbf{Model}                          & \textbf{Metric} & \textbf{Ar} & \textbf{Bn} & \textbf{En} & \textbf{Fi} & \textbf{Id} & \textbf{Ja} & \textbf{Ko} & \textbf{Avg\tnote{$\parallel$}} & \textbf{Ru} & \textbf{Sw} & \textbf{Te} & \textbf{Th} & \textbf{Avg\tnote{$\nparallel$}} & \textbf{Avg} \\ \midrule
\textbf{XLM-R} & Recall      & 0.782                         & 0.797                         & 0.754                    & 0.755                         & 0.840                         & 0.741                         & 0.691   & 0.766                               & 0.741                                  & 0.614                         & 0.820                         & 0.852       &0.757                           & 0.762        \\
\textbf{~~~ + semaCL}                                                                 & Recall          & 0.759       & 0.824       & 0.752       & 0.738       & 0.826       & 0.745       & 0.715       & 0.767                                                                & 0.724       & 0.598       & \textbf{0.851}       & 0.859       & 0.758                                                                 & 0.762                                                        \\   
\textbf{~~~ + langCL (WikiMatrix)}    & MRR             & 0.778       & 0.797        & 0.782       & 0.757       & 0.840       & 0.752       & 0.726       & 0.776                                                                & 0.734       & 0.550       & 0.861       & 0.854       & 0.749                                                                 & 0.766                                                        \\
\textbf{~~~ + langCL (Mr.TyDi)}    & MRR             & 0.767       & 0.842        & 0.739       & 0.749       & 0.819       & 0.761       & 0.713       & 0.770                                                                & 0.724       & 0.605       & 0.773       & 0.857       & 0.739                                                                 & 0.759                                                        \\
\textbf{~~~ + semaCL + langCL (WikiMatrix)} & Recall          & 0.784       & \textbf{0.856}       & \textbf{0.776}       & 0.774       & 0.866       & 0.759       & \textbf{0.738}       & \textbf{0.796}                                                                & \textbf{0.761}       & 0.592       & 0.828       & \textbf{0.887}       & 0.767                                                                 & 0.783                                                        \\
                                                                                                        %  &           &        &        &        &        &        &        &        &        &        &        &        &        &                                                                  &                                                        \\   \midrule
\textbf{~~~ + semaCL + langCL (Mr.TyDi)}    & Recall          & \textbf{0.792}       & 0.806       & 0.768       & \textbf{0.782}       & \textbf{0.871}       & \textbf{0.782}       & 0.737       & 0.795                                                                & 0.755       & \textbf{0.619}       & 0.850        & 0.885       & \textbf{0.777}                                                                 & \textbf{0.786}                                                        \\
                                                                                                        %  &           &        &        &        &        &        &        &        &        &        &        &        &        &                                                                  &                                                        \\  
\bottomrule
\end{tabular}
\begin{tablenotes}
    \item[] Note: Avg for languages with ($\parallel$) and without ($\nparallel$) parallel data.
  \end{tablenotes}
\end{threeparttable}
\vskip -1.5mm
%\caption{Recall@100}
\end{subtable}
\vskip -1mm
\caption{Experiment results when Ru, Sw, Te, Th do NOT have parallel data (Recall@100).}
%\label{tab:IR-nonparallel}
\label{tab:IR-nonpara-Recall}
\end{table*}

\end{document}